\documentstyle[12pt]{article}
\begin{document}
\baselineskip 28pt

\begin{center}
{\bf RARE DECAYS OF HEAVY QUARKS -- SEARCHING GROUND FOR NEW PHYSICS\footnote{Presented
at the XXXIX Cracow School of Theoretical Physics, May 29-June 8, 1999, Zakopane, Poland.}}\\ \ \\
{\bf PAUL SINGER}\\
Department of Physics, Technion--Israel Institute of Technology, Haifa, Israel
\end{center}

The search for new physics beyond the standard model is proceeding nowadays
intensively along experimental and theoretical lines. We review here the
sector of charm radiative decays in this context. The calculation of
$D\rightarrow V\gamma$, $D\rightarrow \ell^+\ell^-\gamma$ transitions
reveals their unequivocal dominance by long-distance contributions.
On the other hand, the beauty-conserving charm-changing electroweak transition
$B_c\rightarrow B_u^*\gamma$ is shown to have unique properties which make it a 
promising avenue in the search for new physics. We describe a calculation of
short- and long-distance contributions to this decay which finds them to be of comparable
size. The branching ratio of this decay in the standard model is estimated to be
$\simeq 10^{-8}$.\\ 

\pagebreak

\begin{center}
{\bf 1. Introduction}
\end{center}

The standard model (SM) of strong and electroweak interactions [1], based on local gauge
invariance with respect to the gauge group SU(3)$_C \times {\rm SU}(2)_L \times {\rm U}(1)_Y$,
is presently in excellent shape. All experimental data are in agreement with SM and 
its relentless
success is providing an ever increasing challenge to both experimentalists and theorists.
The only missing fixture of the model is the SM Higgs boson, for which the existing experimental
searches put a lower bound on its mass of 88.6 GeV/c$^2$ [2].

Despite its remarkable success, it is generally believed that the SM is in fact an effective
theory at the energies presently accessible. This belief is fueled by the fact that the
SM has about 20 arbitrary parameters and, moreover, there is no satisfactory explanation for
many of its salient features. For instance, why does the gauge group of interactions have
the structure expressed by its three factors? Why three generations of fermions?
Why left-right asymmetry? How to explain the observed spectrum of quark and lepton masses
and the pattern of mixing angles?

This situation has led theorists to propose many paths for the possible extension of the
standard model. I do not plan to go into any detail here on the variety of possibilities,
which was reviewed at many conferences [3,4], and I shall restrict myself to the mention of a
few of the more widely-discussed proposals: supersymmetry [5], especially the ``low-energy''
Minimal Supersymmetric Standard Model [6] which is considered as a most likely possibility at the
Fermi scale, grand-unified theories [7], right-left models ]8], two Higgs doublet models [9], 
flavour changing neutral
Higgs models [10], multiple $Z^o$ bosons [11], and anomalous triple gauge boson couplings [12].

An important feature of the standard model is the flavour symmetry, as the gauge interactions
do not distinguish among the three generations of leptons and quarks. In practice, 
this symmetry is broken,
as it is evident from the pattern of masses and mixing angles of the SM fermions. The Higgs
boson is an agent of flavour symmetry breaking in SM via its Yukawa couplings to fermions. However,
this flavour symmetry breaking is realized in a particular way; the tree-level neutral couplings
of the Higgs boson, as well as those of the photon and of the $Z^o$ boson are all flavour diagonal. The
observed neutral flavour-changing processes on the other hand are rather small, being made possible
in SM by loop graphs only; as such their magnitude is determined by the values of quark
masses and of the CKM matrix. In view of the smallness of flavour-changing neutral-current (FCNC) in the
standard model, FCNC transitions are usually considered to be a fertile ground 
for the search of processes induced
by new physics, which does not automatically suppress such processes. The charm sector
plays a special role in this respect, since as a result of the effectiveness of the GIM 
mechanism in this sector, the short distance SM contributions to certain charm processes
are very small. Accordingly, $D^o-\bar{D}^o$ mixing and rare charm decays have been
singled out [4,13] as attractive candidates for the discovery of new physics effects.

In this lecture we consider the potential of the electroweak penguin transitions
$c\rightarrow u\gamma$ in the search for new physics. To begin I shall review shortly the
SM physics of the single quark transition $Q\rightarrow q\gamma$, then I shall present
the status of short-distance and long-distance contributions in processes driven by
$c\rightarrow u\gamma$ and $c\rightarrow u\ell^+\ell^-$ transitions and finally I shall
describe our recent work [14] which singles out the $B_c\rightarrow B_u^*\gamma$ decay as
a unique tool for the search of effects beyond the standard model.\\

\begin{center}
{\bf 2. The flavour changing $Q\rightarrow q\gamma$ transition}
\end{center}

Flavour-changing photon transitions from a heavy quark $Q$ to a light quark $q$ are induced
by loop diagrams and are a basic feature of the standard model [15], generally recognized
as ``electroweak penguins''. Typical transitions are
$s \rightarrow d\gamma$, $b\rightarrow s(d)\gamma$ for which up-quarks contribute in the loop, and
$c\rightarrow u\gamma$, $t\rightarrow c(u)\gamma$ which are driven by down-quarks in the loop.

The amplitude for such transitions, with the quarks $Q,q$ on the mass-shell is given by [15]
\begin{eqnarray}
&&A_\mu^{(Q\rightarrow q\gamma)} = \frac{eG_F}{4\pi^2\sqrt{2}}
\sum_j V^*_{jQ} V_{jq} \bar{u}(q)
\left[F_{1,j}(k^2)k_\mu k\hspace{-0.20cm}/ - k^2\gamma_\mu \cdot \right. \nonumber \\
&& \left. \frac{1-\gamma_5}{2} + F_{2,j} (k^2) i \sigma_{\mu\nu} k^\nu M_Q \frac{1+\gamma_5}{2}
+ m_q \frac{1-\gamma_5}{2}\right] u(Q)  \ . 
\end{eqnarray}

$F_1, F_2$ are the charge-radius and magnetic form factors respectively and $V_{ab}$ are
CKM matrices; $F_1,F_2$ were first calculated in the electroweak SM by Inami and Lim [16].
The $F_1$ term does not contribute to decays with real photons, however, it is relevant in
leptonic decays like $B\rightarrow X(s)\ell\bar{\ell}$, $K\rightarrow \pi\ell\bar{\ell},$
$D\rightarrow V\ell\bar{\ell}$. In order to compare the calculations of these processes with
experiment, one must complement the electroweak SM calculation by the inclusion of QCD
corrections [17]. In this section we shall mention the $s\rightarrow d\gamma$ and 
$b\rightarrow s\gamma$ transitions and in the next section we turn to the
charm sector in more detail.

The contribution of $s\rightarrow d\gamma$ to various radiative $K$-decays [18,19] and
hyperon decays [20-25] has been studied extensively in the last twenty years. As it turns
out [26] radiative processes which are in the $\sim(10^{-4} - 10^{-7})$ range of branching
ratios like $K^+\rightarrow \pi^+\pi^o\gamma$ , $K^+\rightarrow \pi^+e^+e^-$, 
$\Sigma^+\rightarrow p\gamma$, $\Xi^- {\rightarrow} \Sigma^-\gamma$ 
have both short-distance
and long-distance contributions and the latter are dominant; this prevents a direct
and trustworthy check of the SM or of deviations from it in these decays. In order to 
investigate the short distance $s\rightarrow d"\gamma"$ transition one must turn to very
rare decays [27], like $K^+\rightarrow \pi^+\nu\bar{\nu}$,
$K^o_L\rightarrow \pi^oe^+e^-$, $K_L^o\rightarrow \pi^o \nu\bar{\nu}$. In these, the short-distance
contribution is prominent and the QCD corrections to the decay amplitudes
have been estimated [28]. The most frequent of these is $K^+\rightarrow \pi^+\nu\bar{\nu}$,
which is expected
[28] in SM with a branching ratio Br$(K^+\rightarrow \pi^+\nu\bar{\nu})=(9.1\pm3.8)\times 10^{-11}$.
Recently [29], one event has been detected in this channel, which gives 
Br$(K^+\rightarrow \pi^+\nu\bar{\nu})_{\exp} = \left(4.2^{+9.7}_{-3.5}\right)\times 10^{-10}$.
The other two decays are expected with branching ratios of the order of $10^{-11}$ and one
must wait for the planned experiments in order to find out whether the $s\rightarrow d"\gamma"$
and the box diagrams involved of SM give an accurate picture for these transitions. In the
domain of hyperon radiative decays a similar situation prevails [25]; however, there
might be an exception as it appears [22] that the yet unobserved $\Omega^-\rightarrow\Xi^-\gamma$
decay is affected in a measurable manner [24,26] by the SM single quark $s\rightarrow d\gamma$
transition.

Although the $s\rightarrow d\gamma$ was the first to be investigated with the aim of 
relating it to the observed radiative decays of kaons and hyperons, it is the $b\rightarrow s\gamma$
transition [30] which has been the center of attention during the last dozen years. Since
it was pointed out [31] that the enhancement provided by QCD corrections to $b\rightarrow s\gamma$
(in which the top quark in the loop gives the main contribution) would bring the inclusive
$B\rightarrow X_s\gamma$ and exclusive $B\rightarrow K^*\gamma$ decays into the realm of
observability, a considerable amount of theoretical activity has proceeded alongside the
experimental observation. The CLEO collaboration was the first to measure the inclusive
rate [32] Br$(B\rightarrow X_s\gamma)=(2.32 \pm 0.57 \pm 0.35)\times 10^{-4}$ as well the
exclusive (charged and neutral) decay [33] Br$(B\rightarrow K^*\gamma)=(4.5\pm 1.5\pm 0.9)\times 10^{-5}$.
The theoretical effort has been directed on the one hand towards a best determination of the
QCD corrections to the inclusive process in SM and on the other hand to establishing
the limitations imposed by the observed rate on various ``beyond the standard model''
theories. For typical Refs. on the latter effort see [34]. The latest theoretical
calculations within the SM[35] give Br$(B\rightarrow X_s\gamma)=(3.32\pm 0.30)\times 10^{-4}$
which should be compared with two recent experimental results: the CLEO update giving [36]
Br$(b\rightarrow s\gamma) = [3.15\pm 0.35 \ {\rm (stat)} \ \pm 0.32 \ {\rm (syst)} \pm
0.26 \ {\rm (mod)}] \times 10^{-4}$ which is derived from an analysis of $3.3 \times 10^6$
$B\bar{B}$ pairs and the ALEPH result [37] of $[3.11\pm 0.80 \ {\rm (stat)} \ \pm 0.72
\ {\rm (stat)}]\times 10^{-4}$. Obviously, the agreement with the SM is impressive.

There are two remarks to be made here. Firstly, the conclusion on the excellent agreement
with SM assumes that $LD$ contributions are small, which is indeed the result of many calculations
(approximately 5-10\%) [24]. Secondly, we await for experimental results on the 
complementary process
$b\rightarrow s\ell^-\ell^+$ (including $B\rightarrow K^*\ell^+\ell^-$, 
$B\rightarrow K\ell^+\ell^-)$ which should be compared with SM theoretical expectations
of a branching ratio in the $10^{-6}$ range.

Before turning to the charm sector, we conclude that the study of the $Q\rightarrow q\gamma$,
$Q\rightarrow q\ell^+\ell^-$ transitions in SM is waiting for the measurement of very rare decays
in the strangeness domain, while in the beauty sector, where experiments are available,
the standard model does very well so far.\\

\begin{center}
{\bf 3. Short distance $c\rightarrow u\gamma$ and $c\rightarrow u\ell^+\ell^-$}
\end{center}

The $c\rightarrow u\gamma$ transition is induced by the electroweak penguin with the down
quarks running in the loop. In the absence of QCD corrections this transition is extremely small
as a result of the small masses of the quarks in the loop and the smallness of the CKM factors. The
electroweak SM calculation [38] gives for this strongly GIM 
suppressed transition a branching ratio
of $\sim 10^{-17}$ only. Including the QCD corrections at the leading log approximation [38], 
the $C_7$ Wilson coefficient of the $\sigma_{\mu\nu}$ operator gets the admixture of $C_1,C_2$ 
Wilson coefficients and the amplitude is increased by two orders of magnitude, giving a
branching ratio of about $10^{-12}$. The calculation of the complete two-loop QCD corrections [39]
leads, after using unitarity of CKM, to the folowing effective Lagrangian
\begin{eqnarray}
L^{c\rightarrow u\gamma}_{\rm SD} &=&
-\frac{G_F}{\sqrt{2}} \frac{e}{8\pi^2}
V_{cs} V^*_{us} C_7(\mu)\bar{u}\sigma^{\mu\nu}
[m_c(1+\gamma_s) + m_u(1-\gamma_s)] cF_{\mu\nu} \ , \nonumber \\
&&C_7(m_c)=0.0068-0.020 i \ . 
\end{eqnarray}
>From this expression, another increase of two orders of magnitude in the SD amplitude
obtains, giving rise to  $^{\rm SD}\Gamma(c\rightarrow u\gamma)/\Gamma(D^o)\sim 2.5 \times 10^{-8}$.
This implies that in exclusive modes, like $D\rightarrow V\gamma$, the SD contribution
to the branching ratio would be about $(3-5)\times 10^{-9}$. In order to ascertain the
possibility of detecting the SD transition, one must now consider the size of the LD
contribution.

The SD amplitude for $c\rightarrow u\ell^+\ell^-$ can be obtained from the general electroweak
amplitude [16], and the explicit expression for the effective Lagrangian after
certain simplifications is [40]
\begin{eqnarray}
L^{c\rightarrow u\ell^+\ell^-}_{\rm SD} &=&
-\frac{G_F}{\sqrt{2}} \frac{e^2A}{8\pi^2 \sin^2\theta_W}
\bar{u}\gamma^\mu(1-\gamma_5)c\bar{\ell} \gamma_\mu \ell \ ,\nonumber \\
&& A=-0.065   \ . 
\end{eqnarray}
This electroweak transition is not strongly suppressed, in contrast to $c\rightarrow u\gamma$ and although the
QCD corrections have not been evaluated explicitly, they are not expected to change the
value of $A$ appreciably [40]. From (3) one finds $^{\rm SD}\Gamma(c\rightarrow u\ell^+\ell^-)/
\Gamma(D^o)\sim 3 \times 10^{-9}$. Hence, like in the $c\rightarrow u\gamma$ case, one has to ascertain
the LD contribution before one may use these leptonic decays for checking the standard model.\\

\begin{center}
{\bf 4. $D$-mesons radiative decays -- the long distance aspect}
\end{center}

Several treatments have addressed recently the problem of estimating LD contributions to
radiative $D$ decays. These approaches include a pole model [38], a quark model [41] , 
the use of QCD sum rules [42] and 
effective Lagrangians [43]. Already from these works one learns that the $D\rightarrow V\gamma$
decays are expected to have branching ratios of the order of $10^{-4}-10^{-6}$, much larger
than from the SD part.

In a more comprehensive and systematic treatment for these decays [44] we used an effective
hybrid Lagrangian combining heavy quark symmetries and chiral symmetry [45] to calculate nine 
decay modes of the $D\rightarrow V\gamma$ type. The effective nonleptonic Lagrangian used is 
given by
\begin{eqnarray}
L_{LD} &=&
-\frac{G_F}{\sqrt{2}} V_{u q_i}
V^*_{c q_i}
\left[a_1(\bar{u}q_i)^\mu (\bar{q}_jc)_\mu
+ a_2(\bar{u}c)_\mu(\bar{q}_j q_i)^\mu\right] 
\end{eqnarray}
and for the QCD-induced constants $a_1, a_2$ we take $a_1 = 1.26$, $a_2=-0.55$
as determined [46] from nonleptonic $D$ decays. In order to evaluate the matrix elements of
(4) we use the factorization approximation for the $\langle VV_o|(\bar{q_i} q_j)^\mu
(\bar{q}_k c)_\mu|D\rangle$ amplitudes.

The general gauge invariant amplitude for the decay $D(p)\rightarrow V(p_V)+\gamma(k)$ is
\begin{eqnarray}
&&A(D\rightarrow V + \gamma) = \frac{eG_F}{\sqrt{2}} V_{u q_j} \cdot V^*_{cq_j}
\left\{\epsilon_{\mu\nu\alpha\beta} k^\mu\varepsilon^{*\nu}_{(\gamma)} p^\alpha
\varepsilon^{*\beta}_{(V)} A_{PC} \right. \nonumber \\
&&~~~~~ + i\left.\left[(\varepsilon^*_{(V)}\cdot k)(\varepsilon^*_{(\gamma)} \cdot p_{(V)})-
(p_{(V)}\cdot k)(\varepsilon^*_{(V)}\varepsilon^*_{(\gamma)})\right]\right\}
A_{PV} \ . 
\end{eqnarray}
In Ref.~[44] all diagrams contributing to $A_{PC}$, $A_{PV}$ are classified and their explicit expressions
are presented. In Table 1 below we give the predicted widths [44] as well as the existing
experimental upper limits [47]. Since the amplitudes contain several terms, with unknown relative phases, we can present
only their expected range. The first two decays in the Table are Cabibbo-allowed, the
next five are Cabibbo-forbidden and the last two are doubly forbidden. To give an indication,
the photon energy in the first two decays is 717 and 834 MeV respectively. As it is obvious
from Table 1, in all these decays the LD contribution masks totally the SD one -- preventing the 
detection of deviations from it by orders of magnitude.\\

\begin{tabular}{lcc}\hline
$D\rightarrow V\gamma$ Transition & Br Ratio $\times 10^5$ [44] & Exp. limits [47] \\ \hline
& & \\
$D^o\rightarrow \bar{K}^{*o}$~~ & 6-36 & $< 7.6 \times 10^{-4}$ \\
$D^+_s \rightarrow \rho^+$~~~ & 20-80 &  \\ \hline\hline
$D^o\rightarrow \rho^o$~~~~ & 0.1-1 & $< 2.4 \times 10^{-4}$  \\
$D^o\rightarrow \omega$~~~~ & 0.1-0.9 & $< 2.4 \times 10^{-4}$ \\
$D^o\rightarrow \varphi$~~~~ & 0.4-1.9 & $< 1.9 \times 10^{-4}$ \\
$D^+\rightarrow \rho^+$~~ & 0.4-6.3 & \\
$D_s^+\rightarrow K^{*+}$ & 1.2-5.1 & \\ \hline\hline
$D^+\rightarrow K^{*+}$ & 0.03-0.44 &  \\
$D^o\rightarrow K^{*o}$~~ & 0.03-0.2 & 
\end{tabular}

\begin{center}
{\bf Table 1}
\end{center}

Turning now to decays of type $D\rightarrow V\ell^+\ell^-$, these were also calculated
recently [40] using generally the same theoretical framework [45] as for 
$D\rightarrow V\gamma$ transitions. Since the SD transition (Eq.~3) is considerably
larger here than in the $c\rightarrow u\gamma$ case before the application of the QCD corrections, one
could expect that the gap between SD and LD contributions is narrower for the
  leptonic decays in the
SM. Such a situation could open the  window to new physics.

The authors of Ref. [40] have considered the same hadronic transitions as in Table 1. The
SD contribution due to $c\rightarrow u\ell^+\ell^-$ is present in the five Cabibbo suppressed
decays $D^0 \rightarrow (\rho^o,\omega^o, \varphi^o)\ell^+\ell^-$, $D^+\rightarrow \rho^+\gamma$,
$D^+_s\rightarrow K^{*+}\gamma$ while in the other four decays signals for new physics
might come from more exotic contributions. The calculation is performed [40] again using
factorization for matrix elements of (4) which leads to three classes of diagrams: the
annihilation contribution, the $V_o$-spectator part and the $V$-spectator part, where
$V$ is the final state particle and $V_o$ an intermediate vector meson ($\rho,\omega,\varphi)$.
There are thus two kinds of LD contributions: the {\em resonant} mechanism, where in 
addition to $V$ also $V_o$ is produced in the final state and converts to a photon through
vector meson dominance, and a {\em nonresonant} mechanism with the photon emitted directly
from the initial $D$ state, as prescribed by the structure of the hybrid lagrangian [45].
The latter should contain in our approach also possible contributions from intermediate 
$c\bar{c}$
states. The predicted branching ratios for $D\rightarrow V\mu^+\mu^-$, 
including SD + LD contributions,
and the exisiting experimental upper limits are given in Table 2. The range in column two is
due to coupling parameter uncertainties.\\

\begin{tabular}{lcc} \hline
$D\rightarrow V\mu^+\mu^-$ & Calculation [4] of Br(LD+SD) & Exp. limits [48] \\ \hline
& & \\
$D^o \rightarrow \bar{K}^{*o}$ & $(1.6-1.9)\times 10^{-6}$ & $< 1.18 \times 10^{-3}$ \\
$D^+_s\rightarrow \rho^+$ & $(3.0-3.3)\times 10^{-5}$ & \\ \hline \hline
$D^o\rightarrow \rho^o$ & $(3.5-4.7)\times 10^{-7}$ & $< 2.3 \times 10^{-4}$ \\
$D^o\rightarrow \omega^o$ & $(3.3-4.5)\times 10^{-7}$ & $< 8.3 \times 10^{-4}$ \\
$D^o\rightarrow \varphi^o$ & $(6.5-9.0)\times 10^{-8}$ & $< 4.1 \times 10^{-4}$ \\
$D^+\rightarrow \rho^+$ & $(1.5-1.8)\times 10^{-6}$ & $< 5.6 \times 10^{-4}$ \\
$D^+_s\rightarrow K^{*+}$ & $(5.0-7.0)\times 10^{-7}$ & $< 1.4 \times 10^{-3}$ \\ \hline\hline
$D^+\rightarrow K^{*+}$ & $(3.1-3.7)\times 10^{-8}$ & $< 8.5 \times 10^{-4}$ \\
$D^o\rightarrow K^{*o}$ & $(4.4-5.1)\times 10^{-9}$ &  
\end{tabular}

\begin{center}
{\bf Table 2}
\end{center}

The short-distance contributions alone are $\sim 10^{-9}$ for $D^o\rightarrow \rho^o (\omega^o)
\mu^+\mu^-$, $5\times 10^{-9}$ for $D^+\rightarrow \rho^+\mu^+\mu^-$ and 
$1.6 \times 10^{-9}$ for $D^+_s\rightarrow K^{*+} \mu^+\mu^-$, hence
between 2 and 3 orders of magnitude lower than the total Br. The situation is therefore more
favourable than in the $D\rightarrow V\gamma$ case. Branching ratios well above $10^{-6}$
for $D^o\rightarrow (\rho^o,\omega^o)\mu^+\mu^-$ or in the $10^{-5}$ range for 
$D^+\rightarrow \rho^+\mu^+\mu^-$ would be indicative of new physics. It is satisfactory to
note that present experimental bounds are not far above.

Lastly, we mention the $D^{+,o}\rightarrow \pi^{+,o} \ell^+\ell^-$ decays, whose short
distance contribution is again related to $c\rightarrow u\ell^+\ell^-$. In this case,
the LD contribution reaches [49] a branching ratio of the order of $10^{-6}$ in the
$\varphi$-resonance region and a few times $10^{-7}$ in the nonresonant region, a situation
similar to what was encountered in $D\rightarrow V\ell^1\ell^-$ decays.\\

\begin{center}
{\bf 5. $B_c \rightarrow B_u^*\gamma$ --- a unique opportunity}
\end{center}

The situation described in the previous sections indicates that the probability of observing
new physics in $D\rightarrow V\gamma$, $D\rightarrow V\ell^+\ell^-$ or $D\rightarrow P\ell^+\ell^-$
is rather modest. It would require a mechanism which increases the SD amplitude of
$c\rightarrow u\gamma$ or $c\rightarrow u\ell^+\ell^-$ by at least one or two orders
of magnitude, a rather unlikely though not impossible proposition.

Fajfer, Prelovsek and Singer [14] have turned to the domain of very rare decays and have
proposed the idea of exploring the $c\rightarrow u\gamma$ transition when $c$ is embedded
in a beauty particle. In other words, they consider a ``beauty-conserving'' and ``charm-changing''
decay, which is driven by the
$c\rightarrow u\gamma$ transition. As it has been shown by these authors explicitly, such
a transition has about equal SD and LD contributions, making it an ideal testing ground
for deviations from SM [14,50].

The $B_c$-meson, a compact bound state of two heavy quarks of different flavour, $c$ and
$\bar{b}$, has been discovered recently at Fermilab [51] and its lifetime has been determined
as $\tau(B_c)= 0.46^{+0.18}_{-0.16} \pm 0.03$ps. The transition $c\rightarrow u+\gamma$ would
lead to the decay $B_c\rightarrow B^*_u+\gamma$, in which the $\bar{b}$-quark is merely a 
spectator. In order to estimate the SD and the LD contributions to the decay one uses the
effective Lagrangians of (2) and (4). In (2), the appropriate scale for $C_7(\mu)$
is indeed $\mu=m_c$ also for the decay $B_c\rightarrow B_u^*+\gamma$, and not $m_b$, in view
of the spectator role of the $\bar{b}$-quark. The general form of the decay amplitude is as
given in Eq.~(5) and we turn now to the calculation of $A_{PC}$ and $A_{PV}$, which
have both SD and LD contributions.\\
\pagebreak

\begin{center}
{\bf 6. A model for $B_c\rightarrow B_u^*\gamma$}
\end{center}

The SD contribution calculated from (2) can be expressed in terms of two form factors
$F_1(0)$, $F_2(0)$:
\begin{eqnarray}
\varepsilon^*_\mu\langle B^*_u(p',\varepsilon')
|\bar{u} i \sigma^{\mu\nu}q_\nu c|B_c(p)\rangle_{q^2=0} = i \epsilon^{\mu\nu\alpha\beta}
\varepsilon^*_\mu\varepsilon^{'*}_\nu p'_\alpha p_\beta F_1(0) \ , 
\end{eqnarray}

\begin{eqnarray}
&& \varepsilon^*_\mu\langle B^*_u(p',\varepsilon')
|\bar{u} i \sigma^{\mu\nu}q_\nu\gamma_5c|B_c(p)\rangle_{q^2=0} = 
\left[ (M^2_{B_c} - M^2_{B^*_u}) \varepsilon^* \cdot \varepsilon^{'*} \right. \nonumber \\
&&~~~~~~~~~~~~ \left. -2(\varepsilon^{'*} \cdot q) (p\cdot \varepsilon^*)\right] F_2 (0). 
\end{eqnarray}

The LD contributions may be separated into two classes related to the two terms of (5). The
class (I) is related to the $a_2$ term and represents processes $c\rightarrow u\bar{q}_i q_i$
followed by $\bar{q}_iq_i \rightarrow \gamma$, with $\bar{b}$ as spectator.
The $\bar{q}_i q_i \rightarrow \gamma$ transitions are expressed by
$\bar{q}_i q_i$ hadronization into vector meson, thus we have a vector meson dominance
(VMD) approximation. The class II of diagrams is related to the $a_1$ term and corresponds
to the quark process $c\bar{b}\rightarrow u\bar{b}$ with the photon attached to quark
lines. Only the lowest (pole) states are included in the calculation [14,50].

The VMD amplitudes of class I are proportional to
$\varepsilon^*_\mu \langle B^*_u|\bar{u}\gamma^\mu(1-\gamma_5)c|B_c\rangle$ taken at
$q^2=0$. This involves one vector and four axial-vector form factors. However, requirements
of finiteness at $q^2=0$ [46] and gauge invariance imply [14] the vanishing of two axial form
factors and a relation between the other two and accordingly the VMD contribution is expressible in terms 
of two form factors only, $V(0)$ and $A_1(0)$. The amplitudes thus obtained in [14] are
\begin{eqnarray}
A_{PV} &=&-\frac{G_F}{\sqrt{2}}e\left(V_{cs} V^*_{ud}\left[\frac{C_7(m_c)}{2\pi^2}
(m_c - m_u)F_2(0) \right.\right. \nonumber \\
&&\left.\left. + 2 a_2(m_c) C^1_{\rm VMD} \frac{A_1(0)}{M_{B_c} - M_{B^*_u}} \right]\right) 
\end{eqnarray}
\begin{eqnarray}
A_{PC} = &-& \frac{G_F}{\sqrt{2}}e\left(V_{cs} V^*_{ud}\left[\frac{C_7(m_c)}{4\pi^2}
(m_c + m_u)F_1(0) \right.\right.\nonumber \\
&&\left. + 2 a_2(m_c) C^1_{\rm VMD} \frac{V(0)}{M_{B_c} + M_{B^*_u}}\right]
+ V_{cb} V_{ub}^* a_1(m_b) \nonumber \\
&&\left. \times \left[\frac{\mu_{B_c} g_{B^*_c} g_{B^*_u} }{M^2_{B^*_c} - M^2_{B^*_u}} +
\frac{\mu_{B_u} M^2_{B_c} f_{B_c} f_{B_u} } {M^2_{B_c} - M^2_{B_u} } \right] \right) \ . 
\end{eqnarray}

In these expressions the first term is from SD, the second is the LD VMD contribution and the
third term is the LD pole contribution.
Also,
\begin{eqnarray}
C^1_{\rm VMD} = \frac{g^2_\rho(0)}{2M^2_{\rho}} - \frac{g^2_\omega(0)}{6M^2_\omega}
- \frac{g^2_\varphi(0)}{3 M_\varphi^2} = (-1.2 \pm 1.2) \times 10^{-3}{\rm GeV}^2 
\end{eqnarray}
and $\langle V(q,\epsilon)|V_\mu|0\rangle = g_V(q^2)\epsilon^*_\mu$.
$\mu_i, f_i$ and $g_i$ are couplings related to the axial and vector currents and
are defined in [14].

In order to determine the form factors $A_1(0)$, $V(0)$, $F_1(0)$, $F_2(0)$ and the various
$\mu_i, f_i, g_i$ the authors of Ref.~[14]
have chosen the nonrelativistic constituent Isgur-Score-Grinstein-Wise (ISGW) model [52].
This model is considered to be reliable for a state composed of two heavy quarks; in 
addition, the velocity of $B_u^*$ in the rest frame of $B_c$ is to a good measure
nonrelativistic. In the  ISGW model the quarks of mass $M$ move under the influence of
the effective potential $V(r) = - 4\alpha_s/(3r) + c + br$ with $c=-0.81$ GeV, 
$b=0.18$ GeV$^2$ [53]. The authors [14] use variational solutions of the Schr\"{o}dinger
equation, $\psi(\vec{r}) = \pi^{ -\frac{3}{4}} \beta^{\frac{3}{2}} e^{\frac{-\beta^2r^2}{2}}$
for $S$ state with $\beta$ as variational parameter. Using accepted values for current
quark masses, CKM matrix elements and constituent quark masses, one calculates the SD
and the LD contributions separately, as well as the total branching ratio of $B_c\rightarrow
B_v^*\gamma$. It is found [14]:\\

\begin{tabular}{l||ccc} 
& \multicolumn{1}{c}{$Br^{\rm (SD)}$} & $Br^{\rm (LD)}$ & $Br^{({\rm tot})}$\\ \hline
$B_c\rightarrow B_u^*\gamma$ 
& $4.7\times10^{-9}$ 
& $\left(7.5^{+7.9}_{-4.3}\right)\times 10^{-9}$
& $\left(8.5^{+5.8}_{-2.5} \right) \times 10^{-9}$ \\ \hline
\end{tabular}

\begin{center}
{\bf Table 3}
\end{center}

As evidenced by the results of Table 3, the SD and LD contributions are comparable, which in
principle allows one to probe the $c\rightarrow u\gamma$ transition in $B_c\rightarrow B_u^*\gamma$
decay. Experimental detection of $B_c\rightarrow B_u^*\gamma$ at a branching ratio
well above $10^{-8}$ would clearly indicate a signal for new physics. It is worth
mentioning here that at LHC one expects [50] to produce well above $10^8B_c$ mesons.

Finally, we mention a recent calculation of Aliev and Savci [54] which confirms our
conclusions [14,50]. They calculate the SD contribution to $B_c\rightarrow B_u^*\gamma$
by the use of QCD sum rules and find a value for $F_i(0)$ which leads to an 
SD branching ratio for $B_c\rightarrow B_u^*\gamma$ of $\sim 1.6 \times 10^{-8}$, slightly
higher than presented above, but with the same general conclusions.

\begin{center}
{\bf Summary}
\end{center}

We have reviewed the possibility of using various processes to detect deviations from the 
standard model in the charm sector, using the $c\rightarrow u\gamma$, $c\rightarrow u
\ell^+\ell^-$ transitions. The $D\rightarrow V\gamma$ decays are shown to be dominated
by long distance contributions which usually prevents one from observing deviations
from the standard model short distance ones. The situation is somewhat better in
$D\rightarrow V\ell^+\ell^-$ decays, where the gap between SD and LD is smaller.
Here, branching ratios well above $10^{-6}$ for $D^o\rightarrow\rho^o\mu^+\mu^-$ or 
$D^o\rightarrow \omega^o\mu^+\mu^-$ or in the $10^{-5}$ range for $D^+\rightarrow
\rho^+\mu^+\mu^-$ would indicate new physics. Of particular interest is the novel decay
$B_c\rightarrow B^*_u\gamma$ suggested in Ref.~[14]. In this decay both the SD and LD
contributions to the branching ratio are in the $10^{-8}$ range. The SD contribution is
at its natural value. The LD one is strongly suppressed, as follows:
the Class I VMD contribution is very small as a result of the smallness of $C^1_{\rm VMD}$
(Eq.~10), which represents a cancellation of vector mesons contributions at a level
better than 10\% as a result of GIM and SU(3)$_F$ symmetry; on the other hand, the 
class II pole contributions is also strongly supressed in view of the appearance of the factor
$V_{cb} V^*_{ub}$ in the $c\bar{b}\rightarrow u\bar{b}$ pole diagrams. (In $D$ decays we had the
much bigger $V_{cs} V^*_{us}$ factor, which made the LD pole contributions dominant).
This fortuituous occurrence of SD, LD contributions equality establishes the 
$B_c\rightarrow B_u^*\gamma$ decay mode as an ideal testing ground for physics beyond the
standard model. To conclude, we stress that this decay has a clear signature: the
detection requires the observation of a $B_u$ decay in coincidence with two photons --
a high energy one (985 MeV) and a low energy photon (45 MeV) in the respective
centers of mass of $B_c$ and $B_u^*$.\\

\begin{center}
{\bf REFERENCES}
\end{center}
\begin{enumerate}
\item A textbook on the Standard Model is: J.F. Donoghue, E.Golowich and B.R. Holstein,
{\em Dynamics of the Standard Model}, Cambridge University Press (1992).
\item P. Janot, in Proc. '97 Intern. Europhysics Conf. on High Energy Physics, Jerusalem,
Aug. 1997, D. Lellouch, G. Mikenberg and E. Rabinovici (Eds.), Springer Verlag (1998), p. 212.
\item F. Zwirner, in  Proc. '95 Intern. Europhysics Conf. on High Energy Physics, Bruxelles,
July 1995, J. Lemonne, C. Vander Velde and F. Verbeure (Eds.), World Scientific (1996), p. 943.
\item S. Pakvasa, in Proc. FCNC97 Conference, Santa Monica, Feb. 1997, World Scientific (1998);
hep-ph/9705397.
\item J. Wess and J. Bagger, {\it Supersymmetry and Supergravity}, Princeton Univ. Press (1983).
\item S. Dimopoulos and H. Georgi, {\em Nucl.\ Phys.} B{\bf 193}, 150 (1981).
\item H. Georgi and S.L. Glashow, {\it Phys.\ Rev.\ Lett.} {\bf 32}, 438 (1974).
\item R.N. Mohapatra, {\em Unification and Supersymmetry}, Springer Verlag (1986).
\item H.E. Haber, G.L. Kane and T. Sterling, {\em Nucl.\ Phys.} B {\bf 161}, 493 (1979); 
S.L. Glashow and S. Weinberg, {\em Phys.\ Rev.} D {\bf 15}, 1958 (1977).
\item T.P. Cheng and M. Sher, {\em Phys.\ Rev.} D{\bf 35}, 3484 (1987); M. Sher and
Y. Yuan, {\em Phys.\ Rev.} D {\bf 44}, 1461 (1991).
\item J. Hewett and T. Rizzo, {\em Phys.\ Rep.} {\bf 183}, 193 (1989).
\item K. Hagiwara, K. Hikasa, R.D. Peccei and D. Zeppenfeld, {\em Nucl.\ Phys.} B {\bf 282},
253 (1987).
\item J.L. Hewett, Proc. 28 Intern.\ Conf.\ on High Energy Physics (ICHEP 96), Warsaw, Poland, 1996, Z. Ajduk and A.K.
Wroblewski (Eds.), {World Scientific} (1997), vol. 2, p. 1143.
\item S. Fajfer, S. Prelovsek and P. Singer, {\em Phys.\ Rev.} D {\bf 59}, 114003 (1999).
\item M.K. Gaillard and B.W. Lee, {\em Phys.\ Rev.} D {\bf 10}, 897 (1974).
\item T. Inami and C.S. Lim, {\em Prog.\ Theor.\ Phys.} {\rm 65}, 297 (1981).
\item M.A. Shifman, A.I. Vainshtein and V.I. Zacharov, {\em Phys.\ Rev.} D {\bf 18}, 2583 (1978).
\item A.I. Vainshtein et al., {\em Yad.\ Fiz} {\bf 24}, 820 (1976); F.J. Gilman and M.B. Wise,
{\em Phys.\ Rev.} D {\bf 21}, 3150 (1980); L. Bergstr\"{o}m and P. Singer, {\em Phys.\ Rev.\ Lett.} {\bf
55}, 2633 (1985); {\em Phys.\ Rev.} D {\bf 43}, 1568 (1991).
\item M. McGuigan and A.I. Sanda, {\em Phys.\ Rev.} D {\bf 36}, 1413 (1987); H.-Y. Cheng,
{\em Phys.\ Rev.} D {\bf 49}, 3771 (1994). 
\item F.J. Gilman and M.B. Wise, {\em Phys.\ Rev.} D {\bf 19}, 976 (1979).
\item Ya.I. Kogan and M. Shifman, {\em Sov.\ J.\ Nucl.\ Phys.} {\bf 38}, 628 (1983).
\item L. Bergstr\"{om} and P. Singer, {\em Phys.\ Lett.} {\bf 169B}, 297 (1986).
\item M. Nielsen, L.A. Barreiro, C.O. Escobar and R. Rosenfeld, {\em Phys.\ Rev.}, D {\bf 53}, 3620 (1996).
\item G. Eilam, A. Ioannissian, R.R. Mendel and P. Singer, {\em Phys.\ Rev.}, 
D {\bf 53}, 3629 (1996).
\item J. Lach and P. Zencyzykowski, {\em Int.\ J.\ Mod.\ Phys.} {\bf A10}, 3817 (1995).
\item P. Singer, {Proc.\ Workshop on K-Physics}, Orsay, France (June 1996), L. Iconomidou-Fayard (Ed.), Editions
Frontiers (1997), p. 117.
\item L. Littenberg and G. Valencia, {\em Ann.\ Rev.\ Nucl.\ Part.\ Sci.} {\bf 43}, 729 (1993).
\item G. Buchalla, A.J. Buras and M.E. Lautenbacher, {\em Rev. Mod.\ Phys.} {\bf 68}, 1125 (1996).
\item S. Adler et al., BNL 787 Collab., {\em Phys.\ Rev.\ Lett.} {\bf 79}, 2204 (1997).
\item B.A. Campbell and P.J. O'Donnell, {\em Phys.\ Rev.} {\bf D25}, 1989 (1982).
\item S. Bertolini, F. Borzumati and A. Masiero, {\em Phys.\ Rev.\ Lett.} {\bf 59}, 180 (1987);
N.G. Deshpande, P. Lo, J. Trampetic, G. Eilam and P. Singer, {\em Phys.\ Rev.\ Lett.} {\bf 59}, 183 (1987).
\item M.S. Alan et al., CLEO Collab., {\em Phys.\ Rev.\ Lett.} {\bf 74}, 2885 (1995).
\item R. Ammar et al., CLEO Collab., {\em Phys.\ Rev.\ Lett.} {\bf 71}, 674 (1993).
\item F. Borzumati, {\em Zeit.\ f.\ Physik} C {\bf 63}, 291 (1994); J.H. Hewett, Proc.\
21st SLAC Summer Institute, Stanford, Cal., L. De Porcel and C. Dunwoodie (Eds.), SLAC Report
No. 444 (1994), p. 463; J.H. Hewett, hep-ph/9803370.
\item K. Chetyrkin, M. Misiak and M. Munz, {\em Phys.\ Lett.} B {\bf 400}, 206 (1997); {\bf 425}, 414 (E) (1998);
M. Ciuchini, G. Degrassi, P. Gambio and G.F. Giudice, {\em Nucl.\ Phys.} B {\bf 527}, 21 (1998);
A. Kagan and M. Neubert, {\em Eur.\ Phys.\ J.C.} {\bf 7}, 5 (1999); C. Greub, in Proc. 6th Intern.\
Conf.\ on B-Physics at Hadron Machines - Beauty '99, Bled, Slovenia (June 1999), to be
published.
\item R. Bri\`{e}re, in Proc.\ ICHEP98, Vancouver, Canada (1998).
\item R. Barate et al., ALEPH Collab., {\em Phys.\ Lett.} B {\bf 429},
169 (1998).
\item G. Burdman, E. Golowich, J.L. Hewett and S. Pakvasa, {\em Phys.\ Rev.} {\bf D52},
6383 (1995).
\item C. Greub, J. Hurth, M. Misiak and D. Wyler, {\em Phys.\ Lett.} B {\bf 382}, 415 (1996).
\item S. Fajfer, S. Prelovsek and P. Singer, {\em Phys.\ Rev.} D {\bf 58}, 094038 (1998).
\item H.-Y. Cheng et al., {\em Phys.\ Rev.} D {\bf 51}, 1199 (1995).
\item A. Khodjamirian, G. Stall and D. Wyler, {\em Phys.\ Lett.} B {\bf 358}, 129 (1995).
\item B. Bajc, S. Fajfer and R.J. Oakes, {\em Phys.\ Rev.} D {\bf 51}, 2230 (1995);
{\em Phys.\ Rev.} D {\bf 54}, 5883 (1996).
\item S. Fajfer and P. Singer, {\em Phys.\ Rev.} D {\bf 56}, 4302 (1997); S. Fajfer, S. Prelovsek and P.
Singer, {\em Eur.\ Phys.\ J} C {\bf 6}, 471 (1999).
\item R. Casalbuoni et al., {\em Phys.\ Reports} {\bf 281}, 145 (1997).
\item M. Bauer, B. Stech and M. Wirbel, {\em Z.\ Phys.} C {\bf 34}, 103 (1987).
\item D.M. Asner et al., CLEO Collab., {\em Phys.\ Rev.}, {\bf D59}, 092001 (1998).
\item A. Freyberger et al., CLEO Collab., {\em Phys.\ Rev.\ Lett.} {\bf 76}, 3065 (1996); 
{\bf 77}, 2174 (E) (1996); K. Kodama et al., E653 Collab., {\em Phys.\ Lett.} B {\bf 345}, 82 (1995).
\item P. Singer and D.X. Zhang, {\em Phys.\ Rev.} D {\bf 55}, R1127 (1997).
\item S. Prelovsek, S. Fajfer and P. Singer, hep-ph/9905304.
\item F. Abe et al., CDF Collab., {\em Phys.\ Rev.\ Lett.} {\bf 81}, 2432 (1992).
\item N. Isgur, D. Scora, B. Grinstein and M.B. Wise, {\em Phys.\ Rev.} D {\bf 39}, 799 (1989).
\item D. Scora and N. Isgur, {\em Phys.\ Rev.} D {\bf 52}, 2783 (1995).
\item T.M. Aliev and M. Savci, hep-ph/9908203.
\end{enumerate}

\end{document}